# Inverse design of crystal structures for multicomponent systems

Teng Long*, Yixuan Zhang, Nuno M. Fortunato, Chen Shen, Mian Dai, Hongbin Zhang*

*Correspondence to Teng Long and Hongbin Zhang

E-mail: tenglong@tmm.tu-darmstadt.de/ hongbin.zhang@tu-darmstadt.de

Institute of Materials Science, Technical University of Darmstadt, Otto-Berndt-Straße 3, 64287 Darmstadt, Germany

We developed an inverse design framework enabling automated generation of stable multi-component crystal structures by optimizing the formation energies in the latent space based on reversible crystal graphs with continuous representation. It is demonstrated that 9,160 crystal structures can be generated out of 50,000 crystal graphs, leading to 8,310 distinct cases using a training set of 52,615 crystal structures from Materials Project. Detailed analysis on 15 selected systems reveals that unreported crystal structures below the convex hull can be discovered in 6 material systems. Moreover, the generation efficiency can be further improved by considering extra hypothetical structures in the training. This paves the way to perform inverse design of multicomponent materials with possible multi-objective optimization.

**Keywords:** inverse design; machine learning; generative adversarial network; crystal structures;

**Introduction**

Data-driven materials design has developed rapidly in the last decade, commencing with launching of the Materials Genome Initiative [1,2], with the goal to engineer materials with desired properties (dubbed as inverse design hereafter) [3]. There are three widely applied approaches to achieve inverse design of crystalline materials, *i.e.,* high-throughput calculations, global optimizations, and machine learning generative models [4,5]. The high-throughput approach relies on massive density functional theory (DFT) calculations to evaluate the physical properties from a large number of hypothetical structures [6,7]. This method has been successfully applied to identify Li-ion battery anode materials [8], magnetic materials [9], and topological insulators [10,11]. Meanwhile, the global optimization method generalizes the characteristics of materials as an objective function and then predicts the crystal structures (rather than physical properties) by finding the optimal solutions [12]. This approach is best exemplified by the genetic algorithm as implemented in USPEX and CALYPSO [13,14], which has been used to design superconductors [15], superhard materials [16], and magnetic materials [17]. To better utilize the existing data available from databases such as Materials Project (MP) [18], OQMD [6,19], NOMAD [20], and AFLOWlib [21], the machine learning generative models [4] can be divided into two categories: the probabilistic machine learning model and deep learning generative model. The former one predicts physical properties based on the crystal structures with uncertainty quantification, and uses an acquisition function to select potential structures with the desired properties, which has found structures of NaCl and $Y_2Co_{17}$ with limited training data [22,23]. While the latter one relies on a properly defined continuous latent space (explained in details in the following paragraph) which enables optimization of the physical properties, which has been successfully applied on V-O, Mg-Mg-O and Bi-Se systems [24–26].

Focusing on the deep learning generative model, it is essential to construct a continuous latent space. As first demonstrated by Gómez-Bombarelli *et al.* in the inverse design of molecules [27], properties of the designed molecules can be efficiently optimized in the latent space. Similarly, in order to design crystal structures, a latent space consisting of crystal graphs is first established by Xie *et al.* [28] and further improved by Park *et al.* [29], where the crystal graphs can be used as descriptors to optimize the physical properties such as formation energies, band gaps, and so on. However, without the ability to reconstruct crystal structures, such crystal graphs can only conduct forward prediction but not inverse design. Thereafter, with the refinement of continuous representation (voxel and autoencoder) [24,30–32], crystal graphs are constructed to be reversible, which serve as one-to-one mappings of the crystal structures (via encoding and decoding). Therefore, the resulting continuous latent space enables the inverse design of crystalline structures.

The generation model is another key component of a deep learning generative model, which can be best exemplified by variational autoencoder (VAE) [33] and generative adversarial network (GAN) [34]. VAE is a variant of autoencoder, which assumes a distribution function (*e.g.*, Gaussian) of the crystal graphs in the latent space [35]. For instance, Noh *et al.* developed the iMatGen model, and used it to generate novel $V_xO_y$ crystals [24]. Ren *et al.* also used VAE to design inorganic crystals and find novel structures that do not exist in the training set [36]. Court *et al.* generated binary alloys, ternary perovskites, and Heusler compounds by a VAE model [37]. In contrast, the GAN model trains two competing neural networks (*i.e.*, a generator and a discriminator), where the generator is responsible for generating crystal graphs (aiming to minimize their difference with existing crystal graphs), while the discriminator is responsible for distinguishing them (maximize the difference) [26]. This model does not require prior assumptions of a distribution function to start training, which makes it a more attractive method [34]. For example, Nouira *et al.* developed a CrystalGAN model to design ternary A-B-H phases starting from binary A-H and B-H structures, and applied it to the Ni-Pd-H system [38]. The ZeoGAN scheme developed by Kim *et al.* was applied on zeolites, where Wasserstein GAN (WGAN) was used to generate porous materials [39]. Kim *et al.* also developed the Crystal-WGAN to design new structures in the Mn-Mg-O system [25]. Furthermore, our constrained crystals deep convolutional generative adversarial network (CCDCGAN) model has been applied successfully to design novel thermodynamically stable structures in the Bi-Se system [26].

In this work, we generalized CCDCGAN to the multicomponent systems covering all compositions from the whole periodic table. As shown in Fig. 1, one salient feature of our CCDCGAN is to integrate a formation energy prediction model in latent space as a back propagator for the generator [26]. After training on 52,615 structures in the Materials Project database (please refer to the methods for details), the CCDCGAN model has produced 9,160 crystal structures (of which 8310 are distinct, meaning unreported in the training set) out of 50,000 crystal graphs for 667 systems (meaning element combinations, *e.g.*, Cd-Li, Co-Li-O, etc.) covering 71 elements in the periodic table. Detailed analysis on 15 selected systems indicates that 72% of the newly designed crystal structures have negative formation energies after DFT relaxation, and unreported crystal structures below the convex hulls can be identified for 6 systems. Taking the Co-Hf system as an example, we further investigated how to improve the generation efficiency by including additional hypothetical structures in the training. We believe such an implementation can be further generalized to optimize other physical properties beyond the formation energy [28], so that multi-objective optimization can be achieved.

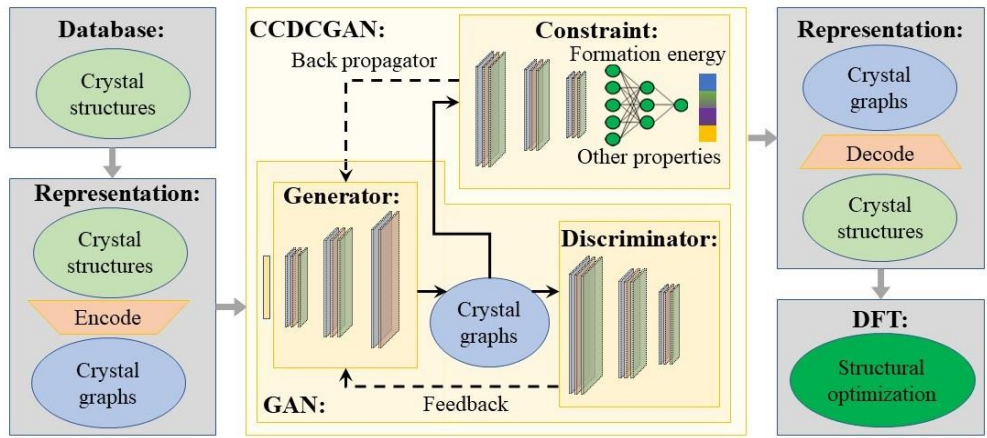

**Fig. 1.** Workflow of CCDCGAN. Three key elements of CCDCGAN, i.e., generator, discriminator, and constraint, are highlighted by shaded blocks in yellow. Such models act on the crystal graphs, which are one-to-one mapping to the crystal structures, with the transformations managed by the autoencoder, *i.e.*, encoder and decoder.

**Methods**

**Data**

Crystal structures and corresponding formation energies from the MP database are used for training in this work, which amount to 125,617 cases (accessed on September 8th 2020). Following Ref. [17] and [20], compounds with less than 20 atoms in primitive cell and with none lattice constant exceeding 10 Å are selected as the training set, leading to 52,615 compounds comprising 84 elements in the periodic table and 25,246 systems. It is straightforward to construct generalized crystal graphs to include compounds with more than 20 atoms or larger dimensions, but the crystal reconstruction ratio will drop significantly because considering larger cells will make it impossible to distinguish subtle crystal structure differences. It is also noted that all the 52,615 compounds are included in the training set for our generative model, which are validated by comparing with the generated structures. Such training set is referred as MP database from here on out.

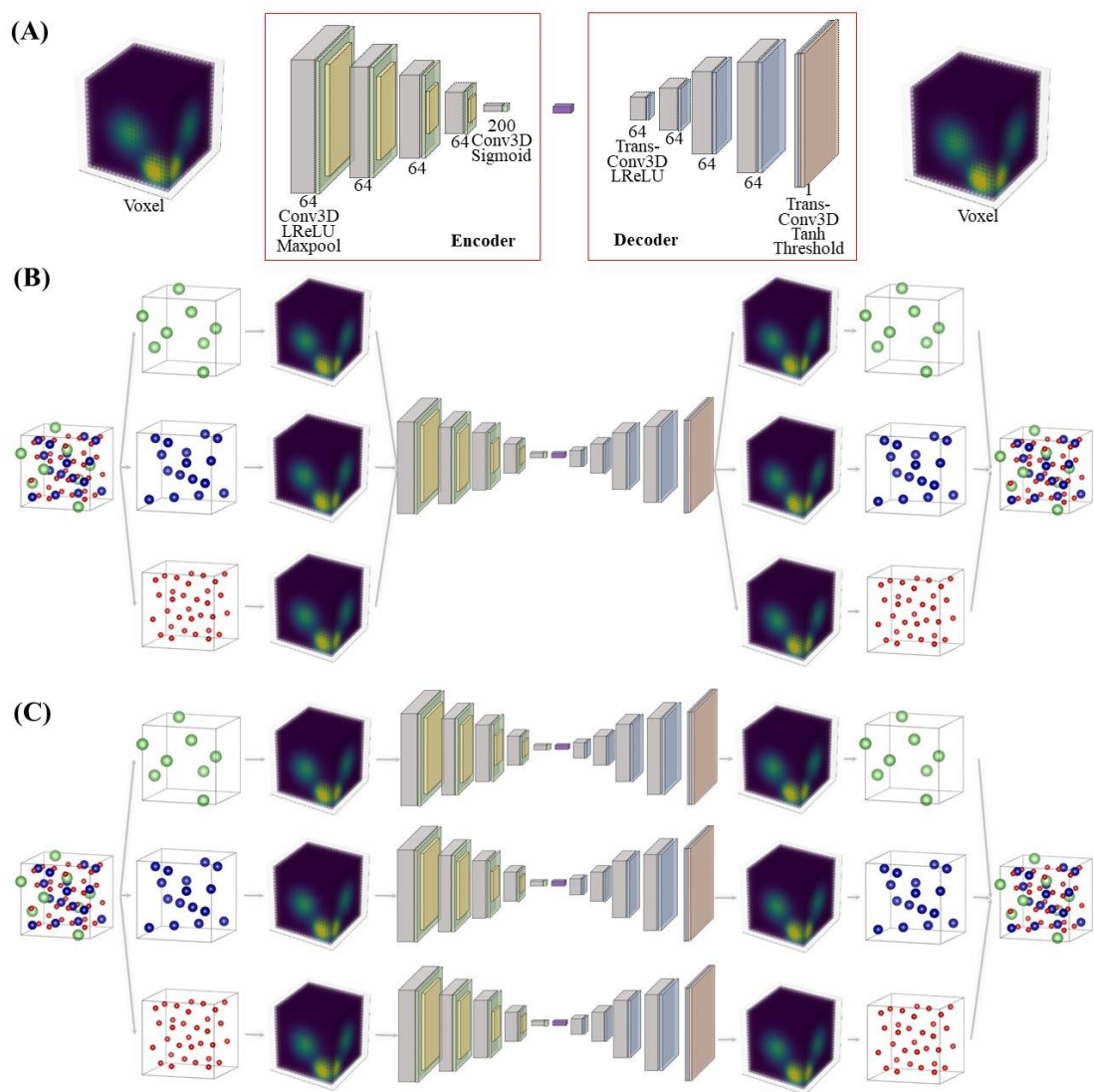

**Fig. 2.** (A) Voxel autoencoder, the autoencoder takes the voxel of crystal structures as input and encodes it into vector in latent space, then the decoder restores the voxel from the encoded vector. (B) Workflow of combined model, different elements share the same autoencoder. (C) workflow of separated model, each element has its own autoencoder.

**Representation**

The crystal graphs are 160×160 matrices, which serve as one-to-one mappings of the crystal structures in the latent space. As shown in Fig. 2(A), such crystal graphs are obtained by encoding 64×64×64 (32×32×32) voxel images for atomic positions (lattice constants) through a 3D convolutional neural network (CNN) typed autoencoder. Importantly, the encoding transformation process can be reversed, *i.e.*, 3D crystal structures can be obtained by decoding the crystal graphs. Correspondingly, the reproduction ratio is an essential indicator on the quality of the crystal graphs. Consistent with Ref. [24] and [26], it is observed that the reproduction ratio can be higher than 90% if only one specific system is considered with sufficient (around 10,000) training structures. However, the encoding-decoding procedure corresponds to Fig. 2(B) can only reach a reproduction ratio of 70% when taking the MP database as input [30]. This can be attributed to the overlapping of different atomic species in the generated crystal graphs. We found that separating the voxels and autoencoders for different elements (as shown in Fig. 2(C)) can reduce the reconstruction loss (Fig. 3(E)) and increase the reproduction ratio to 87%. This leads to reasonable crystal graphs for a large number of multicomponent systems.

**GAN**

To generate new crystal graphs and hence unreported crystal structures, the GAN model is used to sample the latent space, where both the generator and discriminator adopt a CNN model, as shown in Fig. 3 (A) and (B). The generator takes a randomly generated 800-dimensional array obeying the Gaussian distribution as input and generate a 160×160 crystal graph, while the discriminator takes the crystal graph as input and identifies its source. From the mathematical point of view, the goal of the generator (discriminator) is to minimize (maximize) the statistical difference between generated crystal graphs and original crystal structures, which can be expressed for the GAN model as:

$$\max_{D}\left(\min_{G}\left(\frac{1}{2}\cdot E_{\mathbf{x}\sim p_t}\left[1-D(\mathbf{x})\right]+\frac{1}{2}\cdot E_{\mathbf{x}\sim p_g}\left[D(\mathbf{x})\right]\right)\right) \quad (1)$$

where $D$ is the discriminator, $G$ is the generator, $E$ means the expectation value, $\mathbf{x}$ represents the crystal graphs, $D(\mathbf{x})$ is the output of discriminator, *i.e.*, close to 1 (0) means higher probability to be a generated (original) crystal graph, $p_t$ is the possibility density function of the crystal graphs in MP database, while $p_g$ is the possibility density function of generated crystal graphs. In this way, driven by the competition between the generator and discriminator, the GAN model will be competent to design reasonable crystal graphs and thus crystal structures.

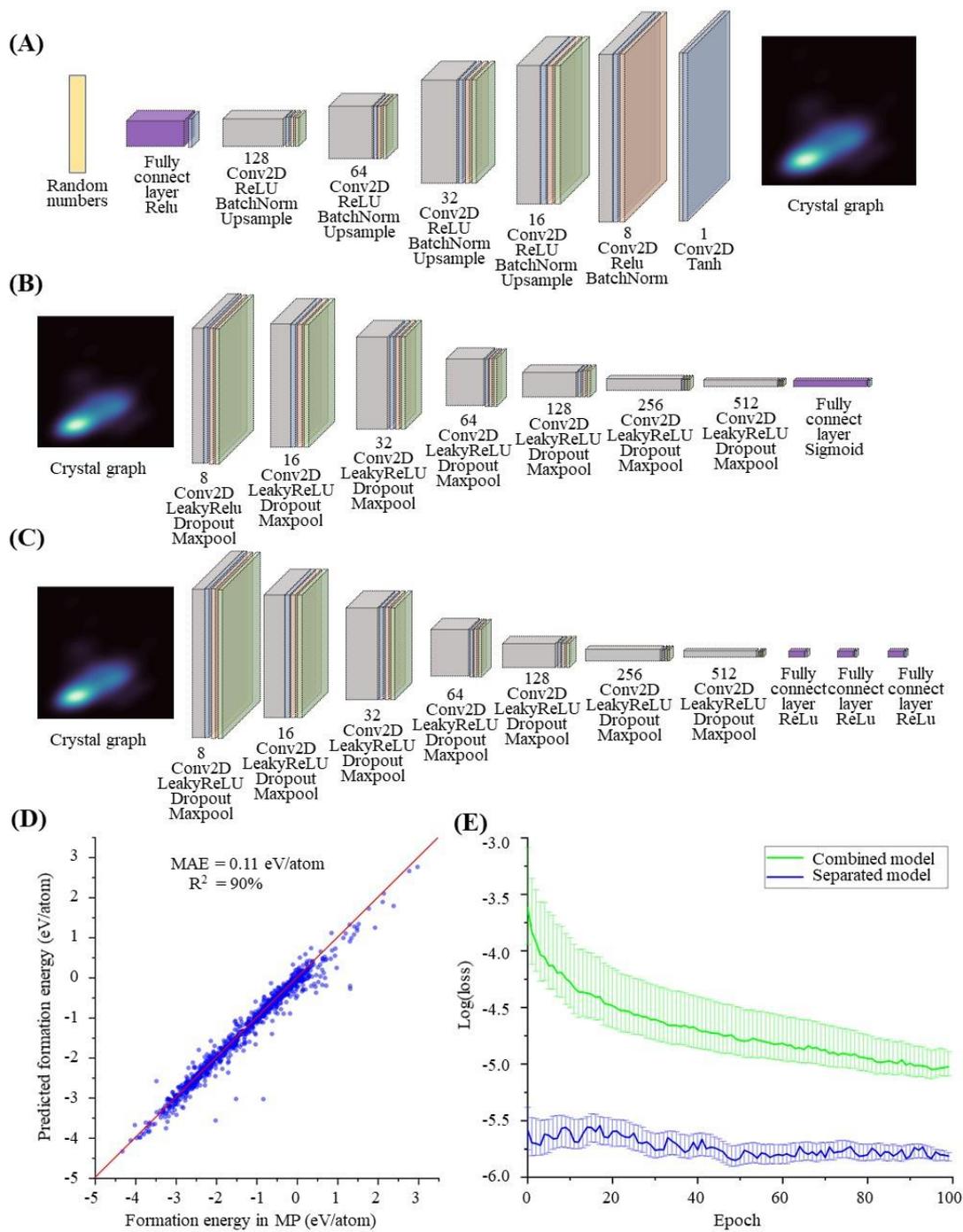

**Fig. 3.** (A) Generator, the generator takes random number as input and generate crystal graph. (B) Discriminator, the discriminator takes the crystal graph as input and judge whether it is generated. (C) Constraint model, the constraint model takes the crystal graph as the input and predict the corresponding properties, *e.g.*, formation energy. (D) Formation energy prediction evaluation by our constraint model. The x-axis is the formation energy in MP database, the y-axis is the formation energy predicted by constraint model. (E) The learning curve of MP database autoencoder, the green line represents test loss of combined model, while the blue line denotes loss of separated model, error bar is the $Q_1$ and $Q_3$ point.

**Constraint**

In order to design crystal structures with desired properties, constraints can be applied to perform optimization in the latent space. In this work, we consider formation energy as the target property so that stable crystal structures can be obtained. Correspondingly, such a constraint can be formulated as a CNN similar to the discriminator, *i.e.*, it takes the crystal graphs as input and predicts their formation energies, as Fig.3(C) demonstrates. For instance, based on the MP database which is divided into a training/validation/test sets following the 80%/10%/10% splitting, the accuracy of predicting formation energies using the crystal graphs as descriptors reaches 90%, as measured by the coefficient of determination ($R^2$), with a corresponding mean absolute error (MAE) of 0.11eV/atom (Fig. 3(D)), which is comparable with the results of CGCNN (Fig. S1)[28].

**CCDCGAN**

Such a constraint on the formation energies has been implemented in our CCDCGAN model as a back propagator coupled to the GAN model, as shown in Fig. 1. That is, the constraint model works also as a feedback mechanism for the generator like the discriminator, but remains unchanged during the CCDCGAN training process, which ensures only the adjustment of generator. It can be mathematically expressed as:

$$\min_{G}\left(\frac{1}{2}\cdot E_{\mathbf{z}\sim p_t}\left[D(\mathbf{z})\right]+\omega\cdot e^{E_f(\mathbf{z})}\right), \mathbf{z}\sim p_g \qquad (2)$$

where $E_f$ is the formation energy predicted by the constraint model, **z** is the generated crystal graph, and $\omega$ is defined as the weight of formation energy loss, which is 0.1 in this model. It is noted that more constraints on the other properties can be integrated so that our CCDCGAN can perform multi-objective optimization, which will be saved for future studies.

**Results and discussion**

Based on our CCDCGAN model trained using the MP database with the formation energies as the constraint, 9,160 crystal structures can be obtained out of 50,000 randomly generated crystal graphs. Such structures comprise 71 elements (Fig. 4(A)), giving rise to 667 systems. And 8,310 structures among them are distinct, indicating variety in structure type. Defining the elemental density as the number of generated structures containing the element divided by the total number of generated structures, the elemental densities of the generated structures are shown in Fig. 4(B), where the elemental densities of the generated structures fluctuated around the density of existed structures, suggesting a similar distribution of elements. Thus, the CCDCGAN model is able to generate a wide range of crystal structures.

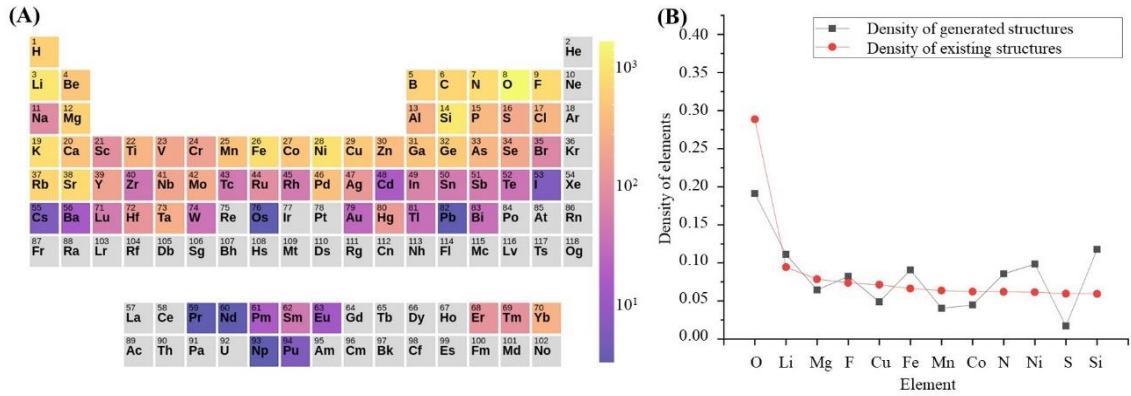

**Fig. 4.** (A) Element distribution of 9,810 generated structures. The Color represents the number of compounds containing the elements, and the color bar is demonstrated in log format. Gray means that this element is not covered by the database. (B) Density of elements of generated structures and existed structures from MP database. The black square represents the density of the generated structures, while the red circle represents the density of the existed structures.

To verify the quality of generated structures, detailed analysis is done on 15 systems focusing on intermetallic compounds (with reasons specified below), including Cr-Ta, Mn-Ta, Fe-Ta, Co-Ta, Ni-Ta, Co-Hf, Al-Cr-Mn, Co-Mn-Si, Cd-Li, Cr-Se, Mn-Se, Fe-Se, Co-Se, Ni-Se, and Co-Li-O, as listed in Table 1. Using the change of formation energies ($\Delta E_f$) of the generated structures before and after DFT optimization as an indicator, it is observed that the average $\Delta E_f$ for 493 generated structures is 0.10 eV/atom, with the largest average $\Delta E_f$ being 0.47 eV/atom for the Cd-Li system. In contrast, $\Delta E_f$ ranges between 0.57 eV/atom and 3.34 eV/atom for structures designed based on the VAE model [37]. In this regard, CCDCGAN is promising in generating nearly optimized crystal structures. Moreover, it is observed that 355 out of 493 (72%) the generated compounds exhibit negative formation energies. That is, compounds in 9 systems have an average formation energy less than zero while those of the other 6 systems are smaller than 70 meV/atom (Table 1). This leads to an average energy above convex hull (referring to the convex hull defined by existing phases) for 6 systems less than 0.1 eV/atom, indicating such generated crystal structures can be synthesized experimentally.

Furthermore, it is observed that distinct structures below the convex hulls can be obtained for 6 out of 15 systems. For instance, the Cd-Li system is an interesting case, where not only all three experimentally known phases ($CdLi_3$, $CdLi$ and $Cd_3Li$) [40] can be reproduced but also two structures (*i.e.*, $Cd_3Li_4$ with a C/2m space group and $Cd_2Li$ with Immm) are predicted, as shown in Fig. 5(A). For other 5 systems, *i.e.*, Mn-Se, Mn-Ta, Ni-Ta, Co-Li-O and Al-Cr-Mn, CCDCGAN generates structures which redefines the convex hull, despite it cannot reproduced all the experimentally known phases. As shown in Fig. 5(B-F), $MnSe_2$ (formation energy -0.47eV/atom, C/2m) and $Ni_2Ta$ (formation energy -0.35eV/atom, I4/mmm) are stable phases that exist experimentally, but are not yet included in the MP database, thus demonstrating the predictive power of CCDCGAN. Additionally, $Mn_4Se_5$ (formation energy -0.41eV/atom, P1), $MnTa_2$ (formation energy -0.17eV/atom, I4/mmm), $AlCr_3Mn$ (formation energy -0.12eV/atom, R3m), $Al_2Cr_3Mn$ (formation energy -0.18eV/atom, Amm2), AlCrMn (formation energy -0.22eV/atom, I4mm), and $AlCr_2Mn_3$ (formation energy -0.11eV/atom, P-3m1) are possible novel stable phases which have not been experimentally reported. Last but not least, two new Co-Li-O phases, *i.e.*, $Co_7LiO_{12}$ (formation energy -1.40eV/atom, P1) and $CoLi_3O_7$ (formation energy -1.32eV/atom, P1), are predicted.

Table 1. Statistical analysis on the generated crystal structures for the 15 selected systems. The second column is number of generated structures (# Gen), while the third column is their average formation energy difference before and after DFT optimization (Mean $\Delta E_f$). The fourth column is the number of structures with $\Delta E_f < 0.2$ eV/atom (# $\Delta E_f < 0.2$) with parenthesis denoting the ratio between # Gen and # $\Delta E_f < 0.2$. The last two column are the average formation energy (Mean $E_f$) and average energy above convex hull (Mean $E_{ab}$), respectively.

| Systems | # Gen | Mena $\Delta E_f$ (eV/atom) | # $\Delta E_f < 0.2$ | Mean $E_f$ (eV/atom) | Mean $E_{ab}$ (eV/atom) |
|---|---|---|---|---|---|
| Cd-Li | 69 | 0.47 | 44 (63.8%) | -0.21 | 0.002 |
| Cr-Se | 11 | 0.02 | 11 (100%) | 0.06 | 0.31 |
| Mn-Se | 11 | 0.03 | 11 (100%) | -0.11 | 0.09 |
| Fe-Se | 36 | 0.25 | 34 (94.4%) | -0.06 | 0.18 |
| Co-Se | 101 | 0.14 | 91 (90.1%) | -0.05 | 0.15 |
| Ni-Se | 36 | 0.12 | 31 (86.1%) | -0.07 | 0.14 |
| Cr-Ta | 4 | 0.13 | 4 (100%) | 0.07 | 0.12 |
| Mn-Ta | 8 | 0.09 | 7 (87.5%) | 0.02 | 0.15 |
| Fe-Ta | 15 | 0.05 | 15 (100%) | 0.01 | 0.11 |
| Co-Ta | 32 | 0.06 | 23 (71.9%) | 0.001 | 0.10 |
| Ni-Ta | 11 | 0.04 | 11 (100%) | -0.07 | 0.12 |
| Al-Cr-Mn | 9 | 0.09 | 8 (88.9%) | -0.04 | 0.10 |
| Co-Li-O | 88 | 0.09 | 82 (93.2%) | -1.18 | 0.07 |
| Co-Mn-Si | 11 | 0.15 | 10 (90.9%) | -0.07 | 0.19 |
| Co-Hf | 51 | 0.18 | 49 (96.1%) | -0.05 | 0.07 |

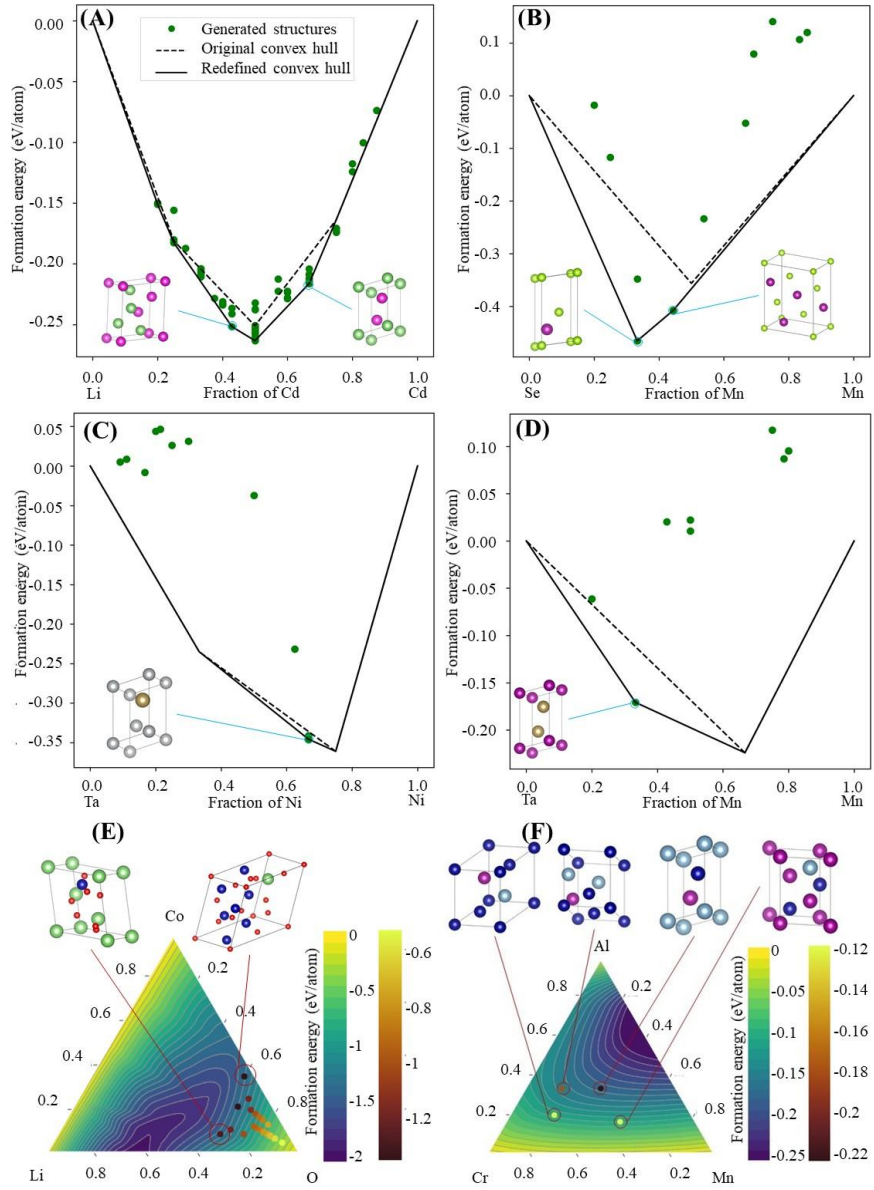

**Fig. 5.** (A) Convex hull of the Cd-Li system, where the red circles denote the structures generated by DCGAN model, green circles indicate the structures generated by CCDCGAN model, dash line and solid line represent the convex hull of MP database and redefined convex hull respectively; crystal structures of $Cd_3Li_4$ and $Cd_2Li$ are demonstrated as well, where green indicates Li atom and magenta denotes Cd atom. (B) Convex hull of the Mn-Se system and crystal structure of $MnSe_2$ and $Mn_4Se_5$, where green represents Se atom and purple means Mn atom. (C) Convex hull of the Ni-Ta system and structure of $Ni_2Ta$, where grey indicates Ni atom and yellow denotes Ta atom. (D) Convex hull of Mn-Ta system and structure of $MnTa_2$. (E) Convex hull of Co-Li-O system, the left one is for the original convex hull and the right one is the redefined convex hull, the color bar represents formation energy; and structures of $Co_7LiO_{12}$, $CoLi_3O_7$, red indicates O atom, blue denotes Co atom. (F) Convex hull of Al-Cr-Mn system, where the cold color triangle represents the convex hull and the warm color circles denote the generated structures, values of the color bar are the corresponding formation energy; the crystal structures of $AlCr_3Mn$, $Al_2Cr_3Mn$, $AlCrMn$, and $AlCr_2Mn_3$ is demonstrated as well, where grey denotes Al atom, blue means Cr atom, magenta indicates Mn atom.

It is noted that for the transition metal oxides, proper calculations should be done based on the DFT+U method to obtain reasonable electronic structure and hence the total energies, due to the strongly correlated nature of the partially filled d-shells. However, the transferability of the U values between compounds with different compositions is limited, which causes possible error in the high-throughput DFT calculations. Besides, additional correction terms are required to evaluate the formation energies and convex hulls of such correlated transition metal oxides with respect to the competing phases [41]. Such a challenge resides in DFT calculations, which is also applicable for rare-earth based systems. Nevertheless, it is beyond the scope of this work, and we suspect our implementation of CCDCGAN is best applicable for predicting novel intermetallic compounds and oxides where mixing DFT+U and DFT calculations is required.

A pending problem is whether CCDCGAN is able to reproduce the experimentally known structures and predict novel structures that redefine the convex hull for the rest 9 systems. Taking the Co-Hf system as an example, all the 51 crystal structures obtained from 50,000 generated cases are above the convex hull, except the CoHf phase is reproduced, as shown in Fig. 6(A). Nevertheless, when the number of generated Co-Hf structures is increased to 2,743, all three known structures (*i.e.*, $CoHf_2$, $CoHf$, $Co_2Hf$) [40] are reproduced, including the $CoHf_2$ phase (formation energy -0.31 eV/atom, Fd-3m) which is not included in the training set. In addition, a $Co_2Hf_5$ (formation energy -0.27 eV/atom, P-1) phase is obtained which lies below the convex hull (Fig. 6(B)). This suggests that CCDCGAN is capable of reproducing the known phases and predicting new crystal structures. Moreover, in order to improve the generation efficiency, we performed explicit DFT calculations on 10,912 hypothetical Co-Hf crystal structures as we did for the Bi-Se system [26], and used these structures as the training set. Such a model can reproduce all three experimental structures and predict $Co_2Hf_5$ phase again after generating only 714 structures, as shown in Fig. 6(C), *i.e.*, 4 times more efficient. Therefore, we believe that increasing the number of compounds in the training set (by taking other database into consideration) will improve the generation efficiency of CCDCGAN significantly.

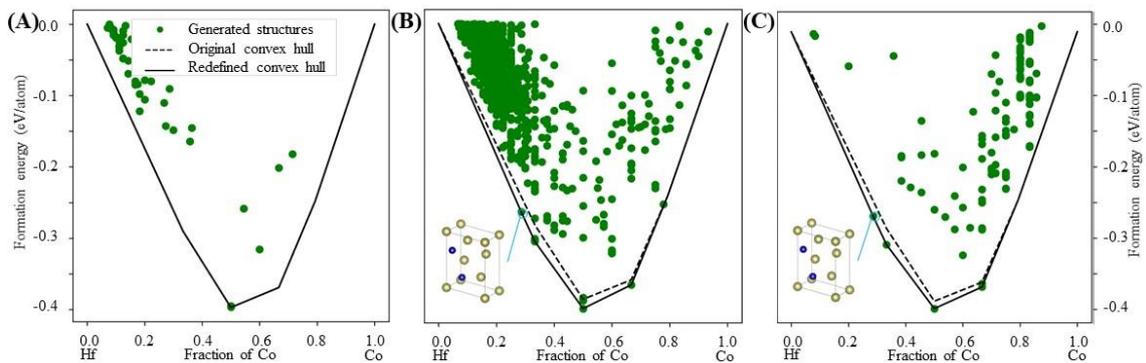

**Fig. 6.** (A) Convex hull of Co-Hf system and the 51 generated structures. (B) Convex hull of Co-Hf system and the 2743 generated structures; the crystal structure of $Co_2Hf_5$, blue means Co atom, yellow is the Hf atom. (C) Convex hull of Co-Hf system and the generated 714 structures after including 10,912 Co-Hf structures in the training set.

## Conclusions

To summarize, we have developed an inverse design framework CCDCGAN to design novel stable crystalline materials by optimizing the formation energies in the latent space defined by crystal graphs, and applied it successfully on multicomponent systems coving most elements in the periodic table. Trained by 52,615 structures in the MP database, a reproduction ratio of as high as 87% is achieved by extending the separated continuous representation model. Correspondingly, CCDCGAN can be used to design distinct crystal structures over a substantial phase space with the generated crystal structures close to their thermodynamic equilibria. Additionally, it is demonstrated that its generation efficiency can be further improved by enlarging the training set by including hypothetical structures. Thus, combined with massive DFT calculations, the CCDCGAN model enables designing novel crystal structures efficiently and hence accelerating the discovery of new materials. In particular, it is speculated that the other physical properties can be optimized in the latent space within the same framework, which will unlock multi-objective optimization to achieve the goal of designing functional materials with optimal performance.

## Acknowledgements

The authors gratefully acknowledge computational time on the Lichtenberg High Performance Supercomputer. Teng Long thanks the financial support from the China Scholarship Council (CSC). Yixuan Zhang thanks the financial support from Fulbright-Cottrell Foundation. Part of this work was supported by the European Research Council (ERC) under the European Union's Horizon 2020 research and innovation programme (Grant No. 743116-project Cool Innov). This work was also supported by the Deutsche Forschungsgemeinschaft (DFG, German Research Foundation) – Project-ID 405553726 – TRR 270.